# Optically modulated field-dependent electrical response in a non-collinear antiferromagnet/ferromagnet heterostructure

Adrielson Dias[1,2], André José[1,2], José Lucas[1,2], Rodrigo Cristian[1,2], and José Holanda[1,2,3,4,5,]*

[1]Group of Optoelectronics and Spintronics, Universidade Federal Rural de Pernambuco, 54518-430, Cabo de Santo Agostinho, Pernambuco, Brazil

[2]Programa de Pós-Graduação em Engenharia Física, Universidade Federal Rural de Pernambuco, 54518-430, Cabo de Santo Agostinho, Pernambuco, Brazil

[3]Unidade Acadêmica do Cabo de Santo Agostinho, Universidade Federal Rural de Pernambuco, 54518-430, Cabo de Santo Agostinho, Pernambuco, Brazil

[4]Programa de Pós-Graduação em Física Aplicada, Universidade Federal Rural de Pernambuco, 52171-900, Recife, Pernambuco, Brazil

[5]Programa de Pós Graduação em Tecnologias Energéticas e Nucleares (Proten), Universidade Federal de Pernambuco, Recife, 50740-545, PE, Brazil

## Abstract

The interplay among optical, electrical, and magnetic stimuli offers opportunities to control the functional response of spintronic heterostructures. Here, we investigate the field-dependent electrical response of $MgO/IrMn_3/Py/Au$ under radio-frequency (RF) excitation, white-light illumination, and direct-current (DC) bias. Under a 20 MHz RF signal with a nominal power of 2 W, the measured voltage in the dark remains at several tens of nanovolts. Upon illumination with approximately 1 W of white light, a pronounced field-dependent feature emerges around 315–320 Oe, reaching 0.9–1.0 μV. The illumination-associated differential voltage is approximately 0.95 μV. A DC current of 1 mA produces a response of approximately 1 μV, while the simultaneous application of RF, DC, and optical excitation increases the voltage to approximately 1.8 μV. Magnetic characterization reveals exchange bias with a blocking temperature near 380 K, while electrical resistance measurements show a pronounced dependence on magnetic-field orientation, consistent with the anisotropic response of the heterostructure. These results demonstrate that the field-dependent electrical response can be modulated through combined optical and electrical excitation, involving magneto-optical interactions and spin-dependent transport phenomena, potentially including spin Hall-related contributions and the magnetic spin Hall effect. This work provides insights into coupled optical, electrical, and magnetic-field effects in heterostructures based on non-collinear antiferromagnets and ferromagnetic layers.

* Corresponding author: joseholanda.silvajunior@ufrpe.br

Orcid iD: https://orcid.org/0000-0002-8823-368X

## 1 Introduction

The integration of magnetic, electrical, and optical functionalities within a single materials platform is a central objective in contemporary spintronics and magneto-optoelectronics. In addition to the control of magnetic order by charge currents, multifunctional heterostructures provide opportunities to modify electronic and magnetic responses through optical excitation, radio-frequency fields, thermal gradients, and interfacial interactions [1–6]. Such control is relevant to magnetic sensing and to device concepts in which more than one external stimulus is used to tune an electrical signal. Antiferromagnets are particularly attractive for spintronic applications because their magnetic sublattices can compensate macroscopically while maintaining robust magnetic order. Their high ordering temperatures, weak stray fields, and unconventional spin-transport properties have stimulated extensive research [3–8]. In non-collinear antiferromagnets, the arrangement of magnetic sublattices can give rise to responses that are not available in conventional collinear systems, including anomalous Hall effects, unconventional spin–orbit torques, and magnetic spin Hall phenomena [9–15].

$IrMn_3$ is a prominent non-collinear antiferromagnet whose triangular spin structure, strong spin–orbit coupling, and compatibility with metallic thin-film heterostructures make it a useful system for studying spin-dependent transport. Its magnetic and interfacial properties depend on crystallographic orientation, chemical order, surface termination, and interface structure [16–20]. When coupled to a ferromagnet such as permalloy (Py), these factors can influence exchange bias, spin transmission, and the electrical response of the bilayer. Previous work has demonstrated magnetic-damping modulation and unconventional spin–orbit effects in $IrMn_3$-based heterostructures [15–17, 21–23]. Optical excitation provides an additional means of perturbing magnetic and electronic systems. Light can modify magnetic order, anisotropy, electronic populations, and interfacial states, although the relevant processes depend strongly on the excitation regime and material system [24–30]. In a conducting multilayer, an illumination-dependent voltage may therefore contain several contributions, including thermal, photoelectric, magneto-optical, and spin-dependent effects.

In this work, we examine the field-dependent electrical response of MgO(100)/$IrMn_3$(100 nm)/Py(10 nm)/Au(3 nm) under 20 MHz RF excitation, white-light illumination, and controlled DC current. We compare dark and illuminated RF measurements, investigate the influence of DC bias, and examine the response under simultaneous excitation. The RF frequency is well below the conventional microwave-frequency range used for ferromagnetic resonance (FMR); accordingly, the measured voltage is treated as a field-dependent electrical response under low-frequency RF excitation, not as evidence of conventional FMR. We define the illumination-associated differential voltage as $V_{M\text{-}O}(H) = V_{RF+Light}(H) - V_{RF+Dark}(H)$. This operational quantity measures the change between illuminated and dark traces under nominally equivalent conditions and is not intended to imply a uniquely identified magneto-optical conversion mechanism.

## 2. Experimental methods

### 2.1. Heterostructure growth and structural characterization

The investigated stack is nominally MgO(100)/$IrMn_3$(100 nm)/Py(10 nm)/Au(3 nm). The MgO(100) single-crystal substrate provides the crystalline template, the $IrMn_3$ layer supplies the non-collinear antiferromagnetic component, the Py layer provides the ferromagnetic component, and the Au cap protects the surface. The $IrMn_3$ films were deposited by magnetron sputtering at a substrate temperature of 570 °C, with a base pressure of approximately $1 \times 10^{-7}$ Torr and an argon working pressure of 3 mTorr. The complete multilayer was prepared by cathodic sputtering. The structural characterization included X-ray diffraction (XRD) and cross-sectional high-resolution transmission electron microscopy (HRTEM). The reported $IrMn_3$ lattice parameter is (0.377 ± 0.001) nm. The diffraction patterns contain the MgO(200) and $IrMn_3$(200) reflections, together with a feature assigned to the Au layer in the complete stack. The HRTEM cross-section resolves the layered arrangement of $IrMn_3$, Py, and Au (**Fig. 1**).

### 2.2. Magnetic and electrical characterization

Magnetic hysteresis measurements were performed using a superconducting quantum interference device (SQUID) magnetometer. Normalized magnetization loops were acquired at 10 and 300 K. The room-temperature measurements followed field cooling from 390 K under a magnetic field of +70 kOe applied along the crystallographic directions [001], [010], and [100]. The exchange-bias field was evaluated from the horizontal displacement of the hysteresis loops, and its temperature dependence was used to estimate the blocking temperature. The field- and angle-dependent electrical resistance was also investigated. The reported measurements include a low-field sweep between −200 and 200 Oe, high-field measurements up to 85 kOe along [001], [011], and [010], and angular scans at applied fields of 1, 5, 10, 30, 70, and 85 kOe. The resistance variation is denoted by ΔR, with DC current of 1 mA.

### 2.3. Optical, RF, and DC measurement configuration

The measurement arrangement combined magnetic-field control, optical illumination, RF excitation, DC bias, and electrical detection (**Fig. 4**). The optical path comprised a xenon lamp, a linear polarizer, a quarter-wave plate, and an optical chopper. The sample was positioned between the magnetic poles, and a photodiode monitored the optical signal. A signal generator supplied the RF excitation. A Schottky-diode detector and oscilloscopes were used to monitor the electrical response, while a Python-based acquisition system controlled the measurement sequence and recorded the voltage as a function of magnetic field. A sinusoidal RF signal at 20 MHz and a nominal output power of 2 W was applied during magnetic-field sweeps. White-light illumination had an approximate optical power of 1 W and was modulated in alternating 5 s illuminated and 5 s light-blocked intervals. Measurements were performed during both negative and positive field sweeps. DC currents of 0.1 and 1 mA were applied in separate measurements while maintaining the RF excitation. Five configurations were considered: (i) RF excitation in the dark; (ii) RF excitation under illumination; (iii) RF excitation with 0.1 mA DC bias; (iv) RF excitation with 1 mA DC bias;

and (v) simultaneous RF excitation, 1 mA DC bias, and illumination. The illumination-associated differential voltage was calculated as

$$V_{M\text{-}O}(H) = V_{RF+Light}(H) - V_{RF+Dark}(H). \quad (1)$$

Here, $V_{RF+Light}(H)$ and $V_{RF+Dark}(H)$ are the voltages measured with and without illumination, respectively, under otherwise nominally equivalent RF and field conditions. This subtraction quantifies the illumination-associated change but does not separate optical, thermal, electronic, contact-related, or magnetic contributions.

## 3. Results and discussion

### 3.1. Structural properties

**Fig. 1** summarizes the structural characterization of the multilayer. The schematic in **Fig. 1(a)** depicts the MgO/$IrMn_3$/Py/Au architecture and the triangular arrangement of the Mn moments in the non-collinear $IrMn_3$ structure. The three magnetic sublattices are separated by angles of 120°. The XRD patterns in **Figs. 1(b)** and **1(c)** show the crystalline contributions assigned to MgO(200) and $IrMn_3$(200), as well as the diffraction feature associated with Au in the complete heterostructure. The cross-sectional HRTEM image in **Fig. 1(d)** resolves the $IrMn_3$, Py, and Au layers. The measured lattice parameter of 0.377 ± 0.001 nm is consistent with values reported for $IrMn_3$ [19, 20]. Together, these observations support the formation of the intended crystalline antiferromagnet/ferromagnet stack.

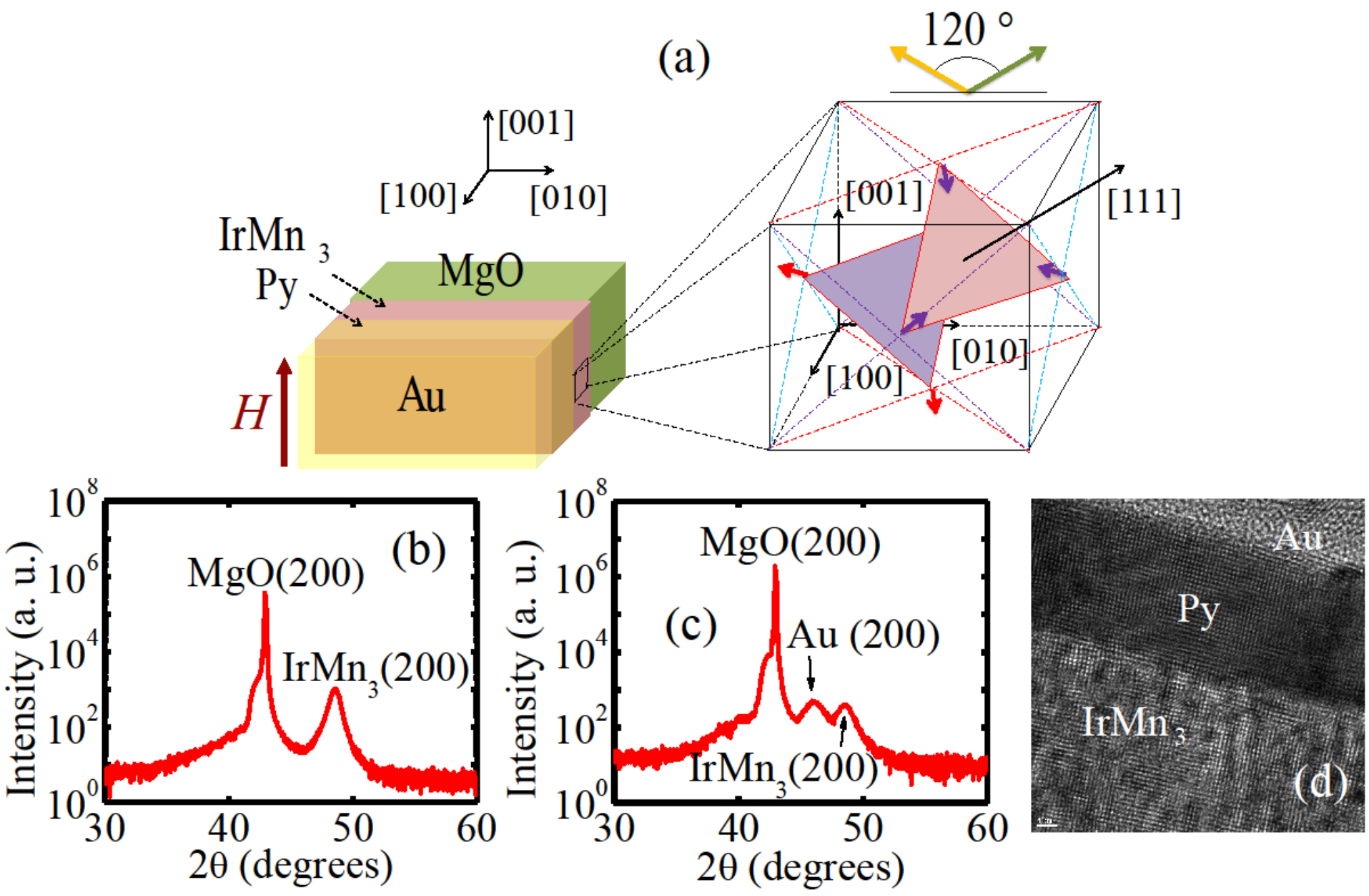

**Figure 1:** Structural characterization of the MgO/$IrMn_3$/Py/Au heterostructure. **(a)** Schematic representation of the multilayer and the non-collinear triangular spin arrangement in $IrMn_3$, with 120° between the magnetic sublattices. **(b)** and **(c)** X-ray diffraction patterns of the MgO/$IrMn_3$ structure and the complete heterostructure, respectively, showing the MgO(200), $IrMn_3$(200), and Au-related diffraction features. **(d)** Cross-sectional HRTEM image resolving the $IrMn_3$, Py, and Au layers.

### 3.2. Exchange bias and magnetic anisotropy

The SQUID hysteresis loops measured at 10 and 300 K are shown in **Figs. 2(a)** and **2(b)**. Both exhibit a horizontal displacement along the field axis, consistent with exchange bias arising from interfacial coupling between antiferromagnetic $IrMn_3$ and ferromagnetic Py. The displacement is larger at 10 K and substantially reduced at 300 K, demonstrating the temperature sensitivity of the interfacial magnetic configuration. The exchange-bias field reaches approximately 62 Oe at low temperature and decreases with increasing temperature, approaching zero near 380 K [**Fig. 2(c)**]. This temperature is identified as the blocking temperature, $T_B$, above which the interfacial antiferromagnetic configuration responsible for the unidirectional anisotropy is no longer stable.

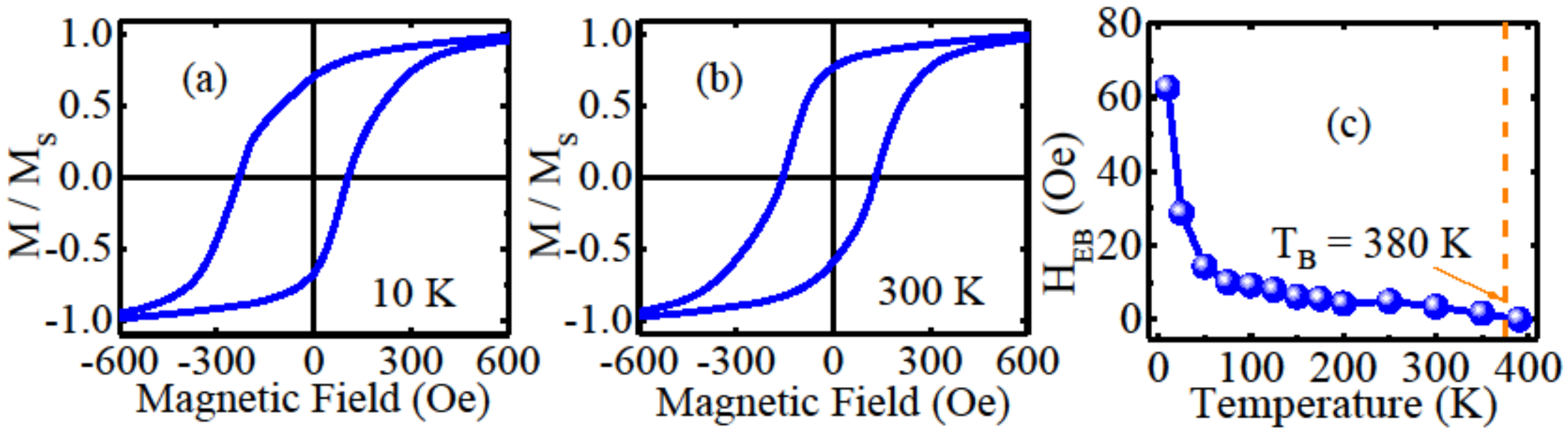

**Figure 2:** Temperature-dependent magnetic hysteresis and exchange bias in the $IrMn_3$/Py heterostructure. **(a)** and **(b)** Normalized magnetization measured at 10 and 300 K, respectively. **(c)** Exchange-bias field as a function of temperature, showing its progressive reduction and disappearance near $T_B \approx 380$ K.

The resistance measurements further reveal a pronounced dependence on magnetic field and orientation (**Fig. 3**). In the low-field range from −200 to 200 Oe, ΔR exhibits a narrow central feature around zero field, with a peak amplitude of approximately 3.5 Ω [**Fig. 3(a)**], with DC current of 1 mA. At high fields, the response differs among the three measured crystallographic directions [**Fig. 3(b)**]: ΔR is positive along [010], negative along [001], and intermediate along [011]. The reported field dependence extends to 85 kOe. Angular scans [**Fig. 3(c)**] display an approximately twofold symmetry. Resistance maxima occur near 90° and 270°, whereas minima are observed near 0°, 180°, and 360°. The modulation amplitude increases with field, from a relatively small variation at 1 kOe to values approaching 3 Ω at higher fields. The curves acquired at 70 and 85 kOe have similar amplitudes, suggesting that the angular response approaches a high-field limiting regime. These observations establish a strong anisotropic component in the electrical transport.

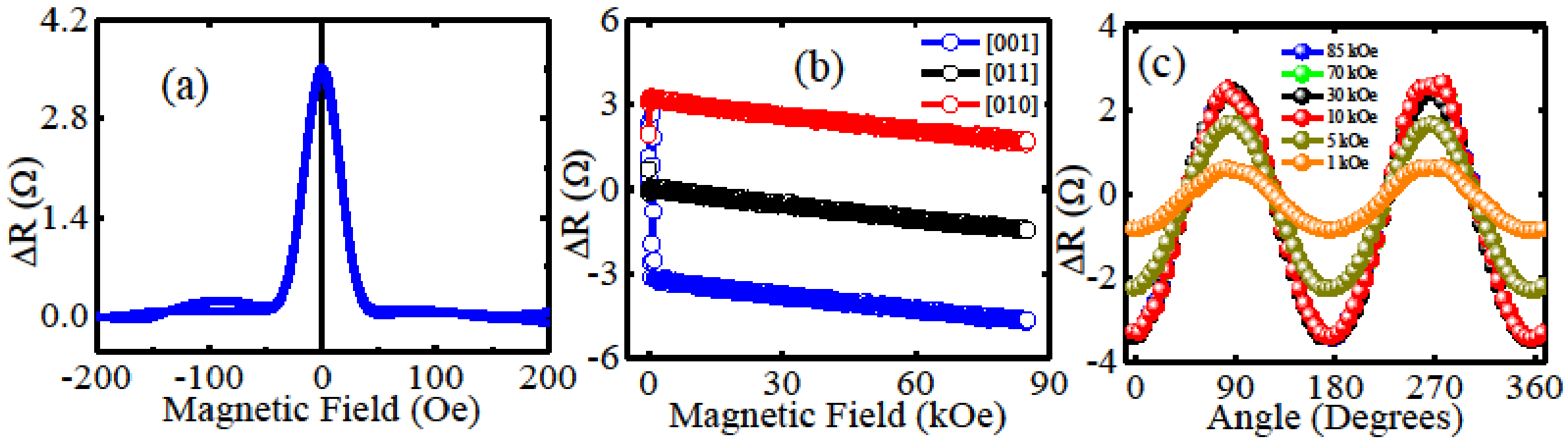

**Figure 3:** Magnetic-field and angular dependence of the electrical resistance variation, ΔR, with DC current of 1 mA. **(a)** Low-field response between −200 and 200 Oe, showing a central feature near zero field. **(b)** ΔR measured along [001], [011], and [010] up to 85 kOe. **(c)** Angular dependence for applied fields of 1, 5, 10, 30, 70, and 85 kOe, showing an approximately twofold symmetry.

The angular dependence presented in **Fig. 3(c)** provides additional evidence of the anisotropic character of the electrical response. The measurements reveal a pronounced twofold symmetry, with resistance maxima occurring approximately at 90° and 270°, and minima around 0°, 180°, and 360°. The amplitude of this angular modulation increases significantly with the applied magnetic field. At 1 kOe, the resistance variation is relatively small, whereas higher magnetic fields produce substantially larger oscillations, reaching approximately 3 Ω. The curves measured at 70 and 85 kOe exhibit similar amplitudes, suggesting that the angular response approaches a high-field saturation regime. Taken together, the results in **Fig. 3** demonstrate a strong dependence of the electrical resistance on magnetic-field magnitude, crystallographic orientation, and angular configuration. The pronounced angular modulation and the distinct responses along the three crystallographic directions highlight the role of magnetic anisotropy in determining the electrical transport behavior. These observations provide further insight into the relationship between magnetic-field-induced changes and anisotropic electrical transport in the investigated system.

### 3.3. Field-dependent response under RF excitation and illumination

The experimental setup integrated magnetic-field control, optical illumination, radio-frequency (RF) excitation, direct-current (DC) bias, and electrical detection. The optical path comprised a xenon lamp, a linear polarizer, a quarter-wave plate, and an optical chopper. The sample was positioned between the magnetic poles, and a photodiode was used to monitor the optical signal. RF excitation was supplied by a signal generator, while a Schottky-diode detector and oscilloscopes were used to monitor the electrical response. A Python-based acquisition system controlled the measurement sequence and recorded the voltage as a function of the applied magnetic field. The optical and electrical excitation conditions were varied to investigate their individual and combined effects on the measured response. The experimental arrangement is shown schematically in **Fig. 4**.

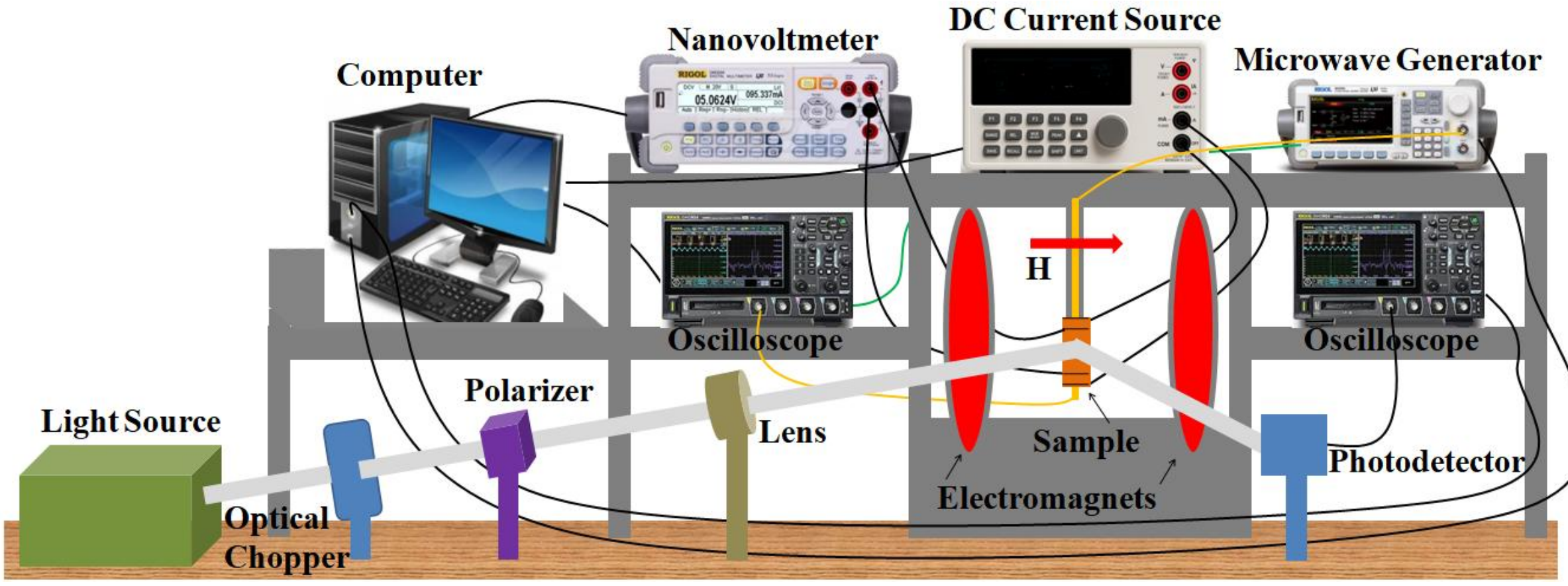

**Figure 4:** Experimental arrangement used to measure the field-dependent electrical response. The optical path comprises a xenon lamp, linear polarizer, quarter-wave plate, and optical chopper. The sample is placed between magnetic poles; a photodiode monitors the illumination. Separate paths provide RF excitation and DC bias, and a computer-controlled acquisition system records the voltage.

The voltage measured under 20 MHz RF excitation in the dark remains at the level of several tens of nanovolts over most of the investigated field range [**Fig. 5(a)**]. No field-localized feature comparable to that observed under illumination is apparent. This dark-state trace provides the reference for assessing the effect of optical excitation. Because the applied frequency is 20 MHz, the data are not interpreted as conventional microwave-frequency FMR. When approximately 1 W of white light is applied during the RF measurement, a pronounced field-dependent feature appears [**Fig. 5(b)**]. The voltage begins to increase near 295–300 Oe, reaches approximately 0.9–1.0 μV around 315–320 Oe, and decreases above approximately 330 Oe. The response is therefore localized in magnetic field rather than appearing solely as a field-independent voltage offset. Continuous white-light illumination can affect a conducting magnetic multilayer through heating, changes in electronic populations, and modifications of magnetic or interfacial properties. The current measurements explore these possibilities. To isolate the difference between the illuminated and dark RF traces, we calculate $V_{M\text{-}O}(H)$ by applying the subtraction defined in Equation (1). The resulting differential curve [**Fig. 5(c)**] exhibits a maximum of approximately 0.95 μV near 315–320 Oe. The signal remains comparatively small below approximately 290 Oe, rises near 295–300 Oe, and decreases above approximately 330 Oe. The persistence of this field-localized feature after subtraction confirms that the illuminated response differs from the dark reference in a field-dependent manner. Here, $V_{M\text{-}O}$ is therefore used strictly as an operational label for the illumination-associated differential voltage.

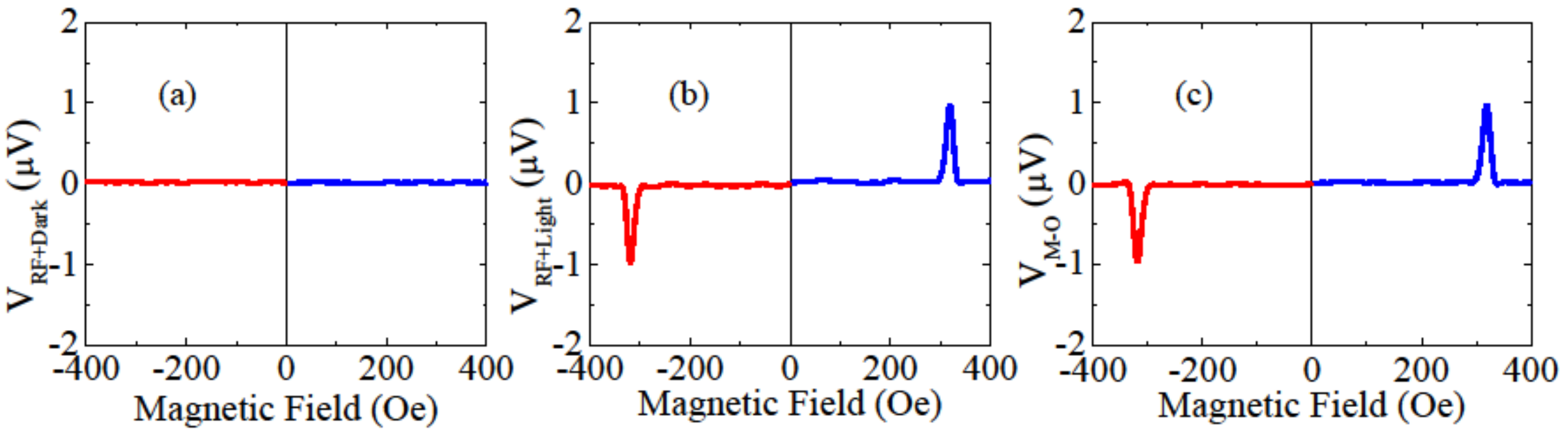

**Figure 5:** Field-dependent electrical response under RF excitation with and without optical illumination. **(a)** Voltage measured in the dark under 20 MHz RF excitation at a nominal output power of 2 W. **(b)** Voltage measured under the same RF conditions with approximately 1 W white-light illumination. **(c)** Illumination-associated differential voltage, $V_{M\text{-}O}(H) = V_{RF+Light}(H) - V_{RF+Dark}(H)$, with a maximum near 315–320 Oe.

### 3.4. Dependence on DC current

The effect of DC bias under continuous RF excitation is summarized in **Fig. 6**. The field-dependent voltage traces acquired at the two DC-current settings are compared in **Figs. 6(a)** and **6(b)**, while **Fig. 6(c)** presents the measured voltage as a function of DC current. Increasing the applied current from 0.1 to 1 mA produces a larger electrical response, reaching approximately 1 μV under the 1 mA condition. Thus, the measured voltage is sensitive to the applied DC bias. The current dependence establishes the mechanism responsible for the enhancement by consolidating the contributions in antiferromagnetic/ferromagnetic heterostructures, including spin-dependent transport, interfacial effects, spin–orbit interactions, and current-induced changes in the magnetic state [9, 10, 17, 23, 31-33].

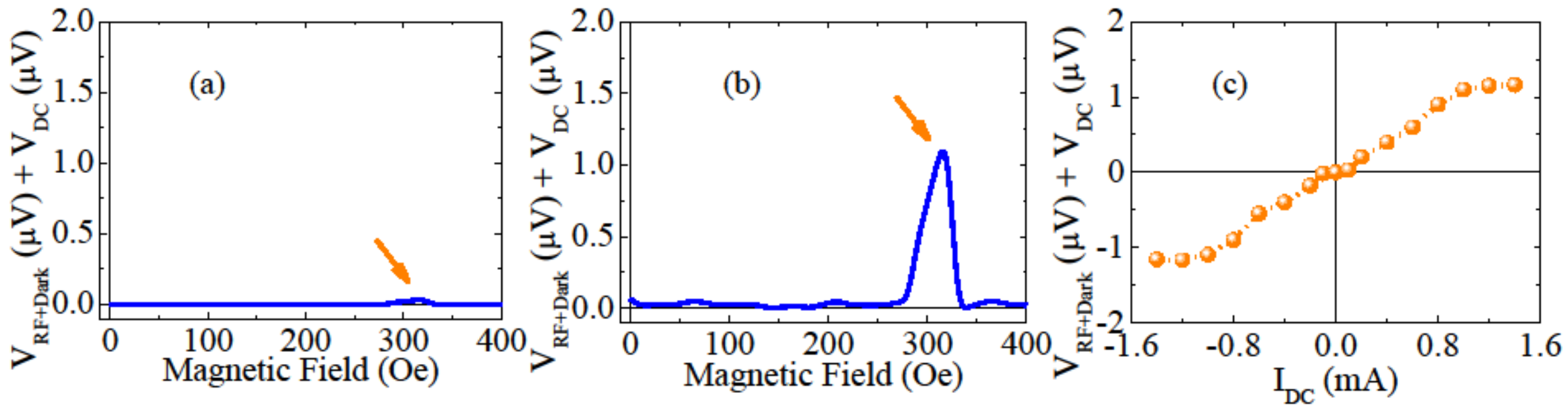

**Figure 6:** DC-current dependence of the electrical response under 20 MHz RF excitation at a nominal output power of 2 W. **(a)** and **(b)** Field-dependent voltage traces acquired at the two DC-current conditions indicated in the panels. **(c)** Measured voltage as a function of applied DC current.

### 3.5. Combined optical, RF, and DC excitation

Under simultaneous application of 20 MHz RF excitation at a nominal output power of 2 W, a DC current of 1 mA, and approximately 1 W white-light illumination, the measured voltage reaches approximately 1.8 μV (**Fig. 7**). This value is larger than the approximately 1 μV response observed under the 1 mA condition and the approximately 0.95 μV maximum illumination-associated differential voltage. The combined excitation therefore produces a measurable response with a greater amplitude than either of these reference values. The comparison demonstrates that the combined response is additive, nonlinear, and governed by magneto-optical interactions between the excitation channels. Thus, the present finding provides evidence of magneto-optical coupling mediated by spin-dependent transport, interfacial effects, spin–orbit interactions, and current-induced changes in the magnetic state.

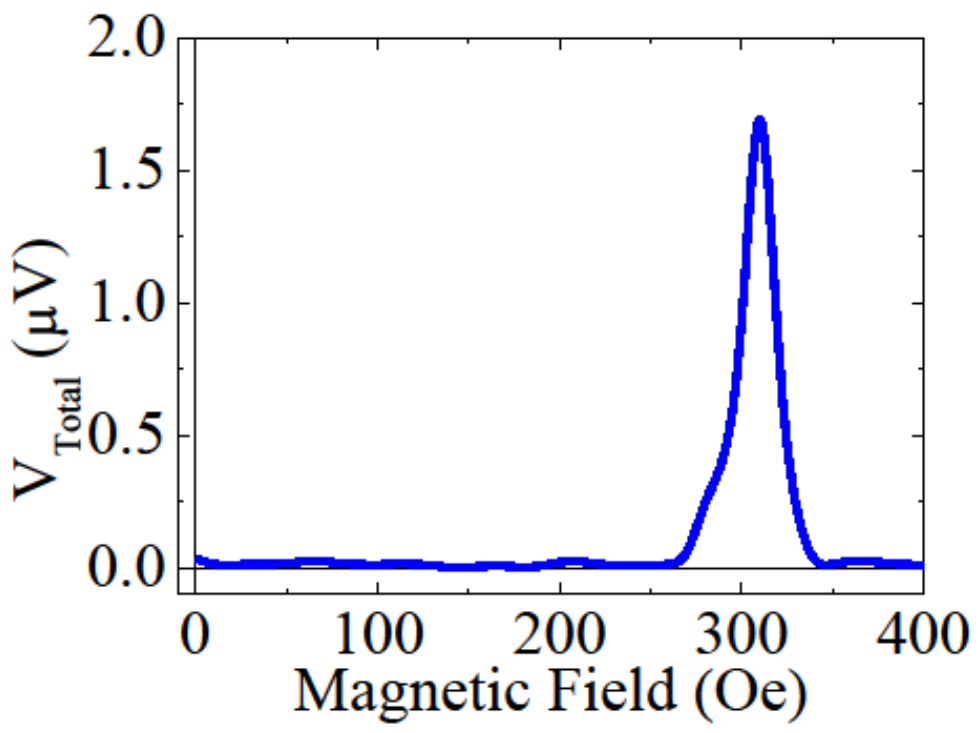

**Figure 7:** Field-dependent electrical response under simultaneous RF excitation, DC bias, and optical illumination. Voltage ($V_{Total}$) measured under 20 MHz RF excitation at a nominal output power of 2 W, a DC current of 1 mA, and approximately 1 W white-light illumination; the response reaches approximately 1.8 μV.

The results establish that the electrical response of the MgO/$IrMn_3$/Py/Au heterostructure depends on magnetic field and can be modified by optical illumination and DC bias in the presence of low-frequency RF excitation. The independent magnetic and resistance measurements show that the sample exhibits exchange bias, a finite blocking temperature, and anisotropic transport. These properties provide a relevant magnetic context for the field-dependent electrical measurements. $IrMn_3$-based heterostructures are known to support orientation-dependent spin Hall conductivity, unconventional spin–orbit torques, and magnetic spin Hall effects [15–17,21–23]. Optical excitation can also influence magnetic and electronic states [24–30]. These precedents motivate investigation of spin-dependent contributions in the present system.

## 4 Conclusions

We have investigated the field-dependent electrical response of MgO(100)/$IrMn_3$(100 nm)/Py(10 nm)/Au(3 nm) under low-frequency RF excitation, optical illumination, and DC bias. The heterostructure exhibits exchange bias that decreases with temperature and vanishes near a blocking temperature of approximately 380 K. Its resistance also displays pronounced magnetic-field and angular anisotropy. Under 20 MHz RF excitation at a nominal power of 2 W, the dark-state voltage remains at the level of several tens of nanovolts. Illumination with approximately 1 W of white light produces a field-localized response near 315–320 Oe, with an illumination-associated differential amplitude of approximately 0.95 μV. A DC current of 1 mA yields a response of approximately 1 μV, while simultaneous RF, DC, and optical excitation produces a voltage of approximately 1.8 μV. These findings demonstrate that the electrical response can be tuned by optical and electrical stimuli in a non-collinear antiferromagnet/ferromagnet heterostructure. The measurements do not establish a single microscopic mechanism, and the 20 MHz excitation is not interpreted as conventional microwave ferromagnetic resonance (FMR). Therefore, this heterostructure provides a platform for future investigations of coupled optical, electrical, and magnetic effects and their relevance to multifunctional spintronic detection.

**Acknowledgements**

This research was supported by the Conselho Nacional de Desenvolvimento Científico e Tecnológico (CNPq), Grant No. 300631/2025-1; the Coordenação de Aperfeiçoamento de Pessoal de Nível Superior (CAPES), Grant No. PROAP2025UFRPE; and the Fundação de Amparo à Ciência e Tecnologia do Estado de Pernambuco (FACEPE), Grant No. APQ-1397-3.04/24.

**Contributions**

A.D., A.J., J.L., and R.C. contributed to the experimental work, data acquisition, and data analysis, and wrote the first version of the manuscript. J.H. revised and supervised the study, coordinated the experimental work, interpreted the results, and finalized the manuscript with contributions from all authors. All authors discussed the results and approved the final manuscript.

**Conflict of interest**

The authors declare that they have no conflict of interest.

**Data Availability Statement**

Data available on request from the authors.